\documentclass[aps,twocolumn,amsmath,showkeys,showpacs,prm]{revtex4-2}
\usepackage{dcolumn}
\usepackage{bm}
\usepackage[T1]{fontenc}
\usepackage{amsmath}
\usepackage{graphicx}
\usepackage{tabularx}
\usepackage{physics}
\usepackage{orcidlink}

\usepackage{hyperref}
\hypersetup{colorlinks=true, linkcolor=blue, citecolor=blue, urlcolor=blue, pdftitle={Spin-phonon Coupling in the Antiperovskite Mn$_3$NiN from First-Principles}, pdfauthor={A. C. Garcia-Castro}}

\begin{document}
\title{Exploration of the spin-phonon coupling in the noncollinear antiferromagnetic antiperovskite Mn$_3$NiN}
 \author{L. Florez-Gomez$^{1,2}$} 
\affiliation{$^1$School of Physics, \textsc{ficomaco} Research Group, Universidad Industrial de Santander, Carrera 27 Calle 09, 680002, Bucaramanga, Colombia.}
\affiliation{$^2$Centro de Investigaci\'on y Estudios Avanzados del IPN - $\textsc{cinvestav}$-Quer\'etaro,  MX-76230, Quer\'etaro, M\'exico.}
\author{W. Ibarra-Hernandez$^{3}$\,\orcidlink{0000-0002-5045-4575}} 
\affiliation{$^3$Facultad de Ingenier\'ia, Benem\'erita Universidad Aut\'onoma de Puebla, Apartado Postal J-39, Puebla, Pue. 72570, M\'exico.}
\author{A. C. Garcia-Castro$^{4}$\,\orcidlink{0000-0003-3379-4495}}
\email{acgarcia@uis.edu.co}
\affiliation{$^4$School of Physics, \textsc{cimbios} Research Group, Universidad Industrial de Santander, Carrera 27 Calle 09, 680002, Bucaramanga, Colombia.}

\begin{abstract} 
Antiferromagnetic antiperovskites, of the form Mn$_3B$N ($B$ = Ni, Cu, Zn, Sn, Ir, and Pt), have shown an outstanding behavior in which, giant negative thermal expansion and chiral magnetic structures are intertwined. As such, aiming to shed light on the magnetostructural behavior, related to the magnetic ordering and structure of this type of materials, we studied by theoretical first-principles calculations the spin-phonon coupling in the Manganese-based antiperovskite Mn$_3$NiN, as a prototype in the Mn$_3B$N family. 
We found a strong spin-lattice coupling by means of the understanding of the phonon-dispersion curves obtained when including the chiral noncollinear magnetic structure. 
To do so, we highlight the importance of the exchange-correlation scheme selected and its influence on the electronic, structural, and vibrational degrees of freedom.
Finally, we found two unstable vibrational modes at the $M$-  and $R$-points when the magnetic ordering is switched to ferromagnetic from the chiral $\Gamma_{4g}$ and $\Gamma_{5g}$. The latter coupling was observed as a key signature of the strong spin-lattice interaction.
\\
\\
DOI:
\end{abstract}


\maketitle

\section{Introduction}
\label{sec1}
Perovskite oxides, and perovskite structures in general ($i.e.$ $ABX_3$), have been in the center of debate and development in the condensed matter field with phenomena ranging from multiferroism and magnetoelectricity \cite{doi:10.1080/00150193.2017.1283171, Fiebig2016, Spaldin2010, Spaldin2019, Spaldin2017} to applications and development in solar cells with hybrid-perovskites \cite{doi:10.1021/acs.chemrev.0c00107, SURESHKUMAR2021940, cite-key}.
More recently, an associated family of materials has caught the attention of the scientific community due to the multifunctional character of its properties that cover, for example, anomalous Hall effect \cite{Gurung2019}, superconductivity \cite{Oudah2016, Oudah2019, Wang2013}, novel batteries \cite{doi:10.1021/ja305709z, D1TA03680G}, ferroelectricity \cite{Garcia-Castro2019}, among other functionalities. 
This family of materials is known as antiperovskite, or inversed-perovskite ($i.e.$ $A_3BX$) \cite{Wang2019, Garcia-Castro2020}, in which, the electrostatic balance and the oxidation site occupation is reversed with respect to the known perovskites \cite{Krivovichev+2008+109+113}. 
In the antiperovskites, the anionic sites occupy the octahedral center, instead of the corners, and the cationic metal transition occupies the corners, instead of the octahedral center, forming the $XA_6$ octahedra.
This change in coordination and atomic occupations give rise to astonishing novel properties.
For instance, when magnetically active cations are placed at the $A$-sites to form the octahedra, added to the triangular geometric coordination of the magnetic sites, a magnetic frustration is induced resulting in chiral noncollinear antiferromagnetic orderings \cite{Fruchart1978}. 
This fascinating magnetic phenomenon is associated with the observed barocaloric response \cite{PhysRevX.8.041035}, large piezomagnetic behavior \cite{doi:10.1021/acsami.8b03112, PhysRevB.78.184414, PhysRevB.96.024451, Gomonaj1989}, topological anomalous Hall effect that can be controlled electrically \cite{Liu2018, Tsai2020}, and spin-torque switching \cite{PhysRevB.101.140405}.
Undoubtedly, the control of all of the latter effects is highly desirable when designing novel and efficient noncollinear spintronics controlled by external electric fields \cite{Qin2019}.
Some efforts have demonstrated the possible control of the magnetic ordering by the application of biaxial strain in thin films \cite{PhysRevMaterials.3.094409}. Moreover, intertwined magnetic and structural transitions are expected to occur as demonstrated by M. Wu \emph{et al.} \cite{Wu2013}.
Additionally, the negative thermal expansion observed in the Mn$_3B$N family \cite{Takenaka2012, Sun2009, Song2011} is a signature of the strong magnetostructural coupling in this class of materials. 
For example, in their perovskite oxides counterparts, the magnetostructural coupling \cite{PhysRevB.84.104440, PhysRevB.85.054417} can induce tangible electric polarization, as in the case of the SrMnO$_3$. Thus, such materials are driven to a multiferroic state thanks to their strong spin-phonon coupling assisted by epitaxial strain \cite{PhysRevLett.104.207204}.  
Consequently, when all the phenomena above are considered, the investigation of the phonon spectra and their coupling with the magnetic orderings can shed light on the physics hidden in these antiferromagnetic antiperovskites.

In this paper, we theoretically study, from first-principles calculations, the magnetostructural coupling by exploring the spin and vibrational degrees of freedom in the Mn$_3$NiN antiferromagnetic antiperovskite, as a prototype among its family. 
Moreover, we performed a careful analysis of the influence of the exchange-correlation on-site Coulomb $U$ term into the physics behind the spin-phonon coupling and the electronic structure in this antiperovskite.
This paper is organized as follows: In Section \ref{secII} are presented all the computational details and the theoretical approaches used in the analysis of this work. 
In Section \ref{secIII} are shown the obtained results and the associated analysis related to the vibrational landscape of the Mn$_3$NiN noncollinear antiferromagnetic antiperovskite. 
Finally, in Section \ref{conclusions} we highlight our conclusions and perspectives in regards to the strong magnetostructural response observed in Mn$_3$NiN.

\section{Computational Details}
\label{secII}
We performed first-principles calculations within the density-functional theory (DFT) \cite{PhysRev.136.B864,PhysRev.140.A1133} approach by using the \textsc{vasp} code (version 5.4.4) \cite{Kresse1996,Kresse1999}. 
The projected-augmented waves, PAW \cite{Blochl1994, PhysRevB.59.1758} scheme was used to represent the valence and core electrons.
The electronic configurations considered in the pseudo-potentials as valence electrons are Mn: (3$p^6$3$d^5$4$s^2$, version 02Aug2007), Ni: (3$p^6$3$d^8$4$s^2$, version 06Sep2000), and N: (2$s^2$2$p^3$, version 08Apr2002). 
The exchange-correlation, $E_{xc}$, was represented within the generalized gradient approximation, GGA-PBEsol parametrization \cite{Perdew2008}, and the $E_{xc}$ of the $d$-electrons was corrected through the DFT$+U$ approximation within the Liechtenstein formalism \cite{Liechtenstein1995}. 
We used a Coulomb on-site value of $U$ = 2.0 eV parameter. The latter optimized to reproduce the experimentally observed  lattice parameter. 
Also, a metaGGA formalism \cite{PhysRevB.84.035117}, within the SCAN implementation \cite{PhysRevLett.115.036402}, was adopted to correlate with the Hubbard correction within the PBEsol$+U$ calculations.
The dependence of the lattice parameter on the electronic correlations was also explored considering the LDA \cite{PhysRevLett.45.566} and PBE \cite{PhysRevLett.77.3865, doi:10.1063/1.478401} exchange correlations.
The periodic solution of the crystal was represented by using Bloch states with a Monkhorst-Pack \cite{PhysRevB.13.5188} \emph{k}-point mesh of 12$\times$12$\times$12 and 800 eV energy cut-off to give forces convergence of less than 0.001 eV$\cdot$\r{A}$^{-1}$ and an error in the energy less than 0.5 meV.  
The spin-orbit coupling (SOC) was included to consider noncollinear magnetic configurations \cite{Hobbs2000}.  
The phonon calculations were performed within the finite-differences methodology \cite{PhysRevLett.48.406, PhysRevB.34.5065} and analyzed through the \textsc{phonopy} interface \cite{phonopy}. 
The modes symmetry was studied with the support of the \textsc{amplimodes} toolkit \cite{Orobengoa:ks5225}. Finally, the atomic structure figures were elaborated with the \textsc{vesta} code \cite{vesta}.


\begin{figure*}[]
 \centering
 \includegraphics[width=16.0cm,keepaspectratio=true]{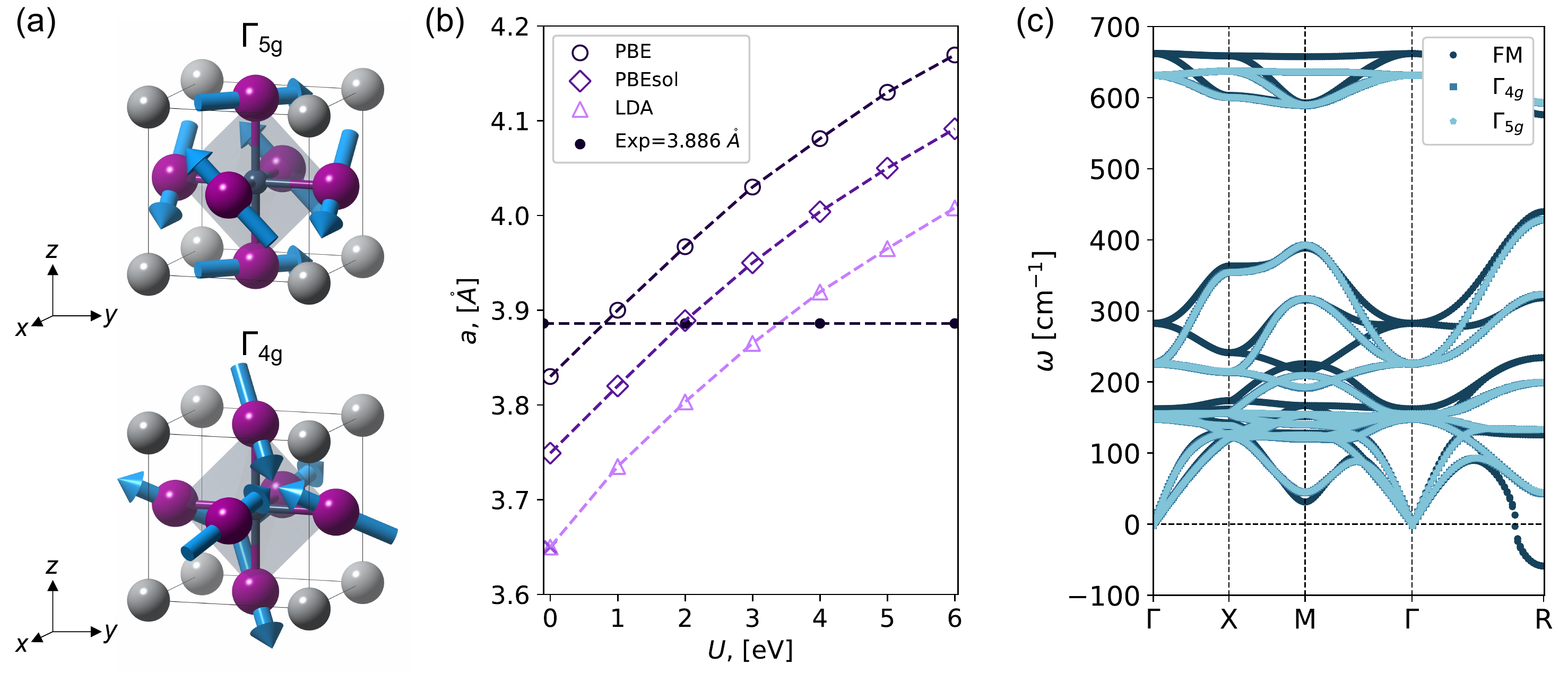}
 \caption{(Color online) In (a) is presented the Mn$_3$NiN $Pm\bar{3}m$ structure in which, the $\Gamma_{5g}$ and $\Gamma_{4g}$ noncollinear ordering are shown. (b) Lattice parameter as a function of the Hubbard $U$ parameter computed for PBE, PBEsol, and LDA pseudopotentials, respectively. The experimentally included lattice parameter was measured at T$_N$ = 262 K below the magnetic transition temperature \cite{Wu2013}. (c) Phonon dispersion curves were obtained for the Mn$_3$NiN at a fixed $U$ value of 2.0 eV and the FM, $\Gamma_{4g}$, and $\Gamma_{5g}$ orderings. The unstable modes are shown at negative frequency values by notation.}
 \label{F1}
\end{figure*}


\section{Results and Discussion}
\label{secIII}
As commented before, in the antiferromagnetic antiperovskites where magnetically active (for example Mn and Ni) cations are placed in the octahedral corner sites, a magnetic frustration is produced as a response of the triangular coordination resulting in the octahedral faces. 
Thus, from the structural point of view, the [111] family of planes form structural Kagome lattices that couples to the resulting magnetic structure after the magnetic frustration is resolved into the symmetry allowed magnetic orderings ($i.e.$ $\Gamma_{4g}$ and $\Gamma_{5g}$ in our case \cite{Fruchart1978}, see Fig. \ref{F1}(a)). 
Therefore, by choosing the $\langle$111$\rangle$ direction as the symmetry axis, the resulting noncollinear structure lies in the [111] plane. 
Then, a strong magnetostructural coupling is expected to be observed in the vibrational landscape of the Mn$_3$NiN when phonon modes that are entangled to the structural Kagome lattices are involved. The latter interaction is mediated by the magnetic exchange coupling constants and Dzyaloshiinski-Moriya interaction that in turn, depends on the bond's length and bonding angles \cite{Dzyaloshinsky,Moriya}.

As a first step into our analysis, we considered the volume effects, widely reported in the literature \cite{doi:10.1088/1468-6996/15/1/015009,Wu2013,PhysRevB.96.024451,Zemen2017}, in prototype antiperovskite materials such as Mn$_3$NiN. 
Moreover, the relationship of the lattice parameter, and associated volume, with the electronic structure in this family of compounds highlights the importance of an appropriate approach to study the spin-phonon coupling (see Fig. \textcolor{blue}{1S} in the Supplement Material).
As such, aiming to properly map the spin-phonon coupling present in this material, we started by exploring the exchange and correlation effects by studying the influence of different pseudopotentials formalisms on the lattice parameter. This structural parameters directly dependent on the electronic RKKY interaction between the Mn-sites forming the octahedra.
Then, we considered LDA, PBE, and PBEsol formalisms in this study. 
We have fixed the magnetic orderings to the noncollinear $\Gamma_{4g}$ and $\Gamma_{5g}$, shown in Fig. \ref{F1}(a), and then, we proceeded with the full optimization of the electronic structure and the lattice parameters.
In both orderings, we obtained, at $U$ = 0.0 eV, lattice parameters of 3.648 \r{A}, 3.830 \r{A}, and 3.749 \r{A} for the LDA, PBE, and PBEsol, respectively, as included in Fig. \ref{F1}(b). 
As expected, the value obtained for PBEsol sits in the middle of LDA and PBE formalisms due to their underestimation and overestimation of the lattice parameters, respectively, acknowledged in the field \cite{Zhang_2018}. 


Experimentally, the observed lattice parameter is $a$ = 3.886 \r{A} at low temperature, below $T_N$ = 262 K \cite{Wu2013} where the frustrated non-collinear magnetism is present. 
Interestingly, although the PBEsol approach is expected to be well-reproduce in comparison to the experimentally observed value, the obtained parameter is well below the expected by $\Delta a$ = 0.137 \r{A}, equivalent to an error of 3.5 \%.
This disagreement is explained in terms of the strong electronic correlation and the entanglement between the magnetic structure and the volume in this type of compounds.
Surprisingly, at the relaxed volume obtained within the LDA approach, the magnetism vanishes and no trace of the magnetic moments is obtained, nor for the Mn-sites, nor for the Ni cell corner sites. 
Moreover, this relaxed lattice parameter is well below the experimentally obtained above the magnetic transition temperature of $a$ = 3.888 \r{A}. In contrast, the net magnetic moments per Mn-atom and noncollinear ordering are obtained when using PBE and PBEsol approaches at the same volume obtained with LDA. 
Here again, suggesting a strong coupling between the magnetic, electronic, and lattice structure.
Therefore, to improve the electronic exchange-correlation misrepresentation, we introduced the DFT+$U$ approach, in which, the $U$ value was considered for the magnetically active Mn-site that is the main responsible for the magnetostructural response.
In Fig. \ref{F1}(b) we present the lattice parameter values, for the LDA, PBE, and PBEsol, obtained as a function of the on-site Coulomb $U$ value. 
We observed a rather large expansion for the three considered schemes. For example, for the PBEsol approach, the lattice parameter varies from 3.749 \r{A} to 4.092 \r{A} when the $U$ value is varied from 0.0 to 6.0 eV. 
This giant expansion of the lattice is due to the strong effect of the electronic structure, and the associated Mn--Mn magnetic exchange, into the structural degrees of freedom possibly influencing the phononic structure.
It is worth noticing that, in the LDA calculation for values above $U$ = 1.0 eV, the expansion and the increase in the correlation correction favor the appearance of magnetism, and then, a development of a net magnetic moment per Mn atom is clearly observed within the noncollinear antiferromagnetic structure. 
The latter is along the lines of the expected negative thermal expansion explained in terms of the existence and annihilation of the frustrated magnetic structure as a function of temperature \cite{Wu2013}.
The observed contraction in the equilibrium volume, in absence of the Hubbard correction, is due to the Mn--Mn overbinding produced by LDA and GGA schemes in metallic Mn-rich compounds \cite{PhysRevB.68.014407, Hobbs_2001, PhysRevB.101.075115}, which in turn, produces a substantial contraction of the lattice parameter, as observed at $U$ = 0.0 eV in Fig. \ref{F1}(b).

After considering a careful characterization of the electronic structure and based-on our previous results aiming to better reproduce the magnetostructural phenomenon and the associated spin-phonon coupling, we have defined the pseudopotential to the PBEsol approach and the on-site correction $U$ value to $U$ = 2.0 eV in the PBEsol+$U$ scheme. 
Moreover, to correlate and corroborate our theoretical strategy with a more advanced approach, we employed the metaGGA formalism within the SCAN representation \cite{PhysRevLett.115.036402}. 
This approach has been successfully applied in the description of metal-transition oxides \cite{PhysRevMaterials.2.095401}, Mn-rich compounds \cite{PhysRevB.101.075115}, and metallic Heusler compounds \cite{PhysRevB.102.045127}.
As such, within the SCAN approach we fully optimized the electronic and structural parameters without the Hubbard correction ($i.e.$ $U$ = 0.0 eV) and conserving a $\Gamma_{4g}$ noncollinear magnetic ordering. 
Under the SCAN representation, we obtained a fully relaxed lattice parameter of $a$ = 3.863 \r{A} representing an error of 0.59 \% with respect to the experimentally obtained value and the computed lattice parameter for PBEsol$+U$, $U$ = 2.0 eV. Additionally, we computed the band-structure for the LDA, PBE, PBEsol, and SCAN representation of the exchange-correlation with and without the $U$ correction, shown in Fig. \textcolor{blue}{2S} in the Supplement Material. 
After comparing the electronic structure with the obtained with the SCAN scheme, it can be observed that the electronic states close to the Fermi energy are considerably improved, for the three schemes, when the on-site Coulomb energy is considered and correctly introduced into the calculations. Nonetheless, the PBEsol$+U$ offers the advantage of being designed to better reproduce the lattice parameter avoiding computational pressure/strain that could alter the electronic structure. 
Then, we conclude that the PBEsol$+U$ methodology fairly reproduces both, the electronic properties and the structural degrees of freedom that might be associated with the vibrational landscape.
Interestingly, most of the reports in the literature \cite{PhysRevB.100.094426,doi:10.1021/acsami.8b03112,PhysRevMaterials.3.094409, Gurung2019} devoted to this type of compounds do not consider the on-site Coulomb $U$ correction in the exchange-correlation and the cell volume is fixed to the experimentally obtained parameter. 
The latter approach might be introducing a fictional negative external pressure, which, in turn, could induce differences in the observed properties due to the strong coupling between the magnetic and the atomic structure. 
Besides, as suggested by Fig. \textcolor{blue}{2S}, the electronic structure strongly differs from the one obtained with more advanced methodologies, such as meta-GGA. 
For this reason, the exchange-correlation correction is needed to better reproduce the electronic structure close to the Fermi energy, and with it, the magnetic response and structural behavior.
Thus, the computational and theoretical methodology proposed here is highly recommended when treating this class of compounds.


\begin{figure*}[]
 \centering
 \includegraphics[width=16.0cm,keepaspectratio=true]{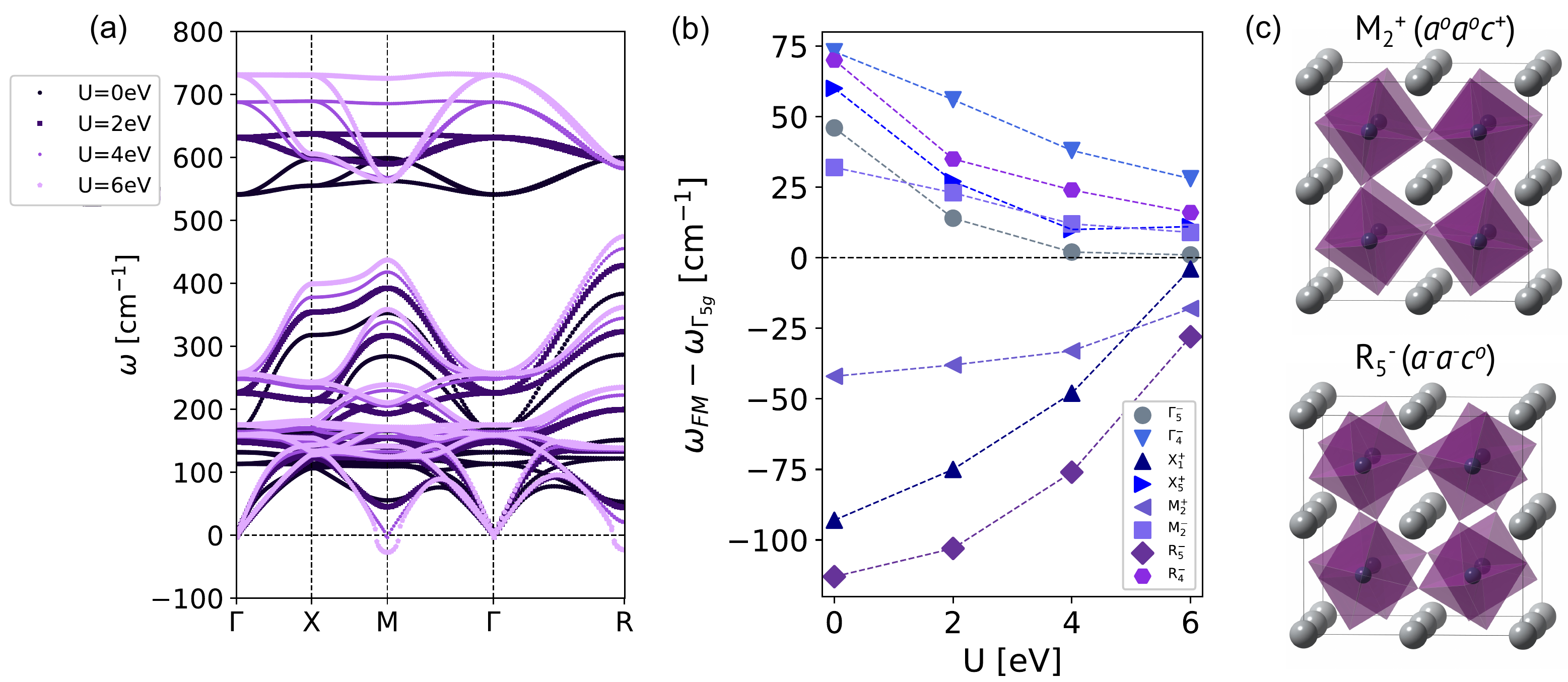}
 \caption{(Color online) (a) Phonon dispersion for the $\Gamma_{5g}$ ordering in the Mn$_3$NiN computed at different $U$ values in which, the lattice volume was fixed to the experimentally obtained value. The latter is in order to explore the effect of the correlation into the vibrational modes. Here, the unstable modes are shown at negative frequency values by notation. (b) Obtained change in the frequencies, $\Delta \omega$ = $\omega_{FM}$ $-$ $\omega_{\Gamma_{5g}}$, when the magnetic transitions in induced. In the latter, the extracted phonon frequencies of the lowest vibrational modes at the high-symmetry points in the Brillouin zone.  (c) Representation of the structural distortions associated with the $M_2^+$ and $R_5^-$ modes where the octahedral rotation and tilting can be appreciated, respectively.}
 \label{F2}
\end{figure*} 


Once the noncollinear magnetic structure is considered into the symmetry operations, the group symmetry is reduced from the $Pm\bar{3}m$ (SG. 221) space group to the $R\bar{3}m$ (SG. 166) although the point group is conserved to the $\bar{3}m$. 
The magnetic symmetry groups are defined as $R\bar{3}m'$ (MSG. 166.101) and $R\bar{3}m$ (MSG. 166.97) for the $\Gamma_{4g}$, and $\Gamma_{5g}$ magnetic orderings, respectively, see Fig. \ref{F1}(a).
Next, after carefully having defined the exchange-correlation and the Hubbard correction effects into the electronic and atomic structure, we proceeded by obtaining the full phonon-dispersion curves for the noncollinear \textsc{fm} (111), $\Gamma_{4g}$, and $\Gamma_{5g}$ magnetic orderings, as shown in Fig. \ref{F1}(c). 
Due to the metallic character of the Mn$_3$NiN antiperovskite and the frustrated magnetic structure that is resolved in the non-collinear chiral magnetic $\Gamma_{4g}$ and $\Gamma_{5g}$ orderings, we adopted the finite-differences method in which, we considered the non-collinear magnetic ordering at each atomic displacements configuration. The latter displacements are obtained by symmetry into the 2$\times$2$\times$2 supercell in which the interatomic forces are computed and later used to reconstruct the dynamical matrix tensor and with it, the full phonon-dispersions.
In Fig. \ref{F1}(c) we present the phonon-dispersion curves for the 3 magnetic orderings at the fully relaxed lattice parameter of $a$ = 3.885 \r{A} with $U$ = 2.0 eV. 
It is relevant mentioning that despite a symmetry lowering is observed when the noncollinear magnetic orderings are introduced, within the margin error of 2.0 cm$^{-1}$, the $Pm\bar{3}m$ and $R\bar{3}m$ phonons are indistinguishable.
For this reason, the high-symmetry points are selected with respect to the $Pm\bar{3}m$ Brillouin zone, BZ.
As it can be appreciated, for the chiral $\Gamma_{4g}$ and $\Gamma_{5g}$ noncollinear magnetic orderings, the phonons and vibrations are fully stable along the entire path in the BZ with non-observable tangible differences in the vibrational landscape between them.
In contrast, for the \textsc{fm} (111) ordering, a negative ($i.e.$ unstable) branch is observed at the $R$-point. This unstable phonon has a frequency value of $\omega$ = $-$59 $i$cm$^{-1}$.
Clearly, this unstable phonon observed, as a result of the magnetic transition between the \textsc{fm} to $\Gamma_{4g}$, or $\Gamma_{5g}$, is proof of the strong spin-lattice coupling present in this antiperovskite.

\begin{table}[t]
\centering
\caption{Phonon frequencies, in cm$^{-1}$, at the high-symmetry points computed for the $\Gamma_{4g}$, $\Gamma_{5g}$, and $\textsc{fm}$ orderings and keeping the on-site Coulomb value of $U$ = 2.0 eV.}
\begin{tabular}{c  c  c  c  c  c  c}
\hline
\hline
Ord. / Ph. mode        &  $\Gamma_5^-$  & $\Gamma_4^-$ & $\Gamma_4^-$   & $\Gamma_4^-$   \rule[-1ex]{0pt}{3.5ex} \\
\hline
$\Gamma_{4g}$ &  149 &  155 & 226   &  631   \\
$\Gamma_{5g}$  &  148 & 155&   226   &  631  \\
$\textsc{fm}$ &  151 & 162 & 282   &  662 & \\
\hline
        & $X_1^+$ & $X_1^+$  & $X_5^+$   & $X_3^-$ &  $X_1^+$ &  $X_5^+$  \rule[-1ex]{0pt}{3.5ex} \\
\hline
$\Gamma_{4g}$ &  127 &  155  & 214 & 354 & 600 & 637 \\
$\Gamma_{5g}$  &  127 &  155 &  214   &  354 & 600 & 637 \\
$\textsc{fm}$ &  131 &  162 &  241   & 363 & 603 & 659\\
\hline
        &  $M_2^+$  & $M_5^-$   & $M_2^-$ &  $M_4^+$ &  $M_5^-$ &  $M_2^-$ \rule[-1ex]{0pt}{3.5ex} \\
\hline
$\Gamma_{4g}$ & 45 &  141   &  193 & 392 & 590 & 636 \\
$\Gamma_{5g}$  &  45 &  140   &  193 & 392 & 590 & 636 \\
$\textsc{fm}$ &  31 &  152   &  216 & 389 & 592 & 657\\
\hline
        &  $R_5^-$  & $R_4^-$   & $R_4^-$ &  $R_3^-$ &  $R_2^-$  & $R_5^+$\rule[-1ex]{0pt}{3.5ex} \\
\hline
$\Gamma_{4g}$ &  43 &  133   &  199 & 324 & 428 & 591 \\
$\Gamma_{5g}$  &  44 &  133   &  199 & 323 & 428 & 592 \\
$\textsc{fm}$ &  $-$59 &  126   &  234 & 319 & 440 & 576 \\
\hline
\hline
\end{tabular}
\label{tab:1}
\end{table}

In Table \ref{tab:1} we present the extracted frequency values for the phonons in the Mn$_3$NiN at the $\Gamma$, $X$, $M$, and $R$ high-symmetry points.
We observed that the frequency values computed for the $\Gamma_{4g}$ and $\Gamma_{5g}$ are almost the same in agreement with the degenerated magnetic states. 
Now, when compared to the \textsc{fm} (111) ordering, we found that several specific modes are highly affected due to the different magnetic orderings. 
This is the case of the $\Gamma_4^-$,  $X_5^+$, $M_2^+$, and $R_5^-$ where the computed |$\Delta \omega$| values are 56 cm$^{-1}$, 27 cm$^{-1}$, 14 cm$^{-1}$, and 103 cm$^{-1}$, respectively.
After analyzing these modes, we observed that the atomic displacements associated with each of them are related to the Mn--Mn interaction responsible for the magnetic response. 
In the $\Gamma_4^-$ case, this mode involves an in-phase octahedral stretching accompanied with a displacement of the nitrogen atoms present in the center of the octahedron. 
The $X_5^+$ mode is related with an in-phase stretching involving only the Mn atoms positioned in the equatorial sites of the octahedron alongside with movements of the nitrogen atoms. 
The case of the $M_2^+$ mode is also interesting because involves a rigid in-phase octahedral rotation. 
The most affected mode, the $R_5^-$, involves a rigid out-of-phase octahedral rotation that conserves the Mn--Mn distance unaffected, as in the $M_2^+$ case, but, alters the bonding angles inducing a distortion within the [111] Kagome lattice.

For our next step in the analysis, we studied only the effect of the electronic correlations in the vibrations and the magnetic structure. Then, we fixed the lattice parameters to the relaxed structure within the PBEsol$+U$ ($U$ = 2.0 eV), and then, we vary the $U$ value from 0.0 to 6.0 eV also keeping the $\Gamma_{5g}$ and \textsc{fm} magnetic orderings fixed. 
In Fig. \ref{F2}(a) we present the phonon-dispersion curves obtained at the fixed volume for $U$ = 0.0, 2.0, 4.0, and 6.0 eV considering the $\Gamma_{5g}$ ordering. 
Despite the fixed volume, we found that the vibrational modes are strongly affected by the exchange-correlation correction where, in general, a hardening of the branches is observed as the on-site Coulomb term is increased.
Interestingly, for values of $U$ = 6.0 eV, there is an appearance of two unstable phonons at the $M$- and $R$-points.
Here, for $U$ values above 4.0 eV, the $M_2^+$ and $R_5^-$  modes become unstable as a consequence of the large on-site Coulomb repulsion that is affecting the exchange coupling, and in turn, modifying the modes'  frequencies.
The latter modes are identified as $M_2^+$ and $R_5^-$ and are representing the in-phase octahedral rotation and the out-of-phase octahedral tilting, respectively,  as shown in Fig. \ref{F2}(c).

In Fig. \ref{F2}(b) we present the differencies, $\Delta \omega$ = $\omega_{\textsc{fm} }$ $-$ $\omega_{\Gamma_{5g}}$, in the phonon frequency, for the most relevant modes close to 0 cm$^{-1}$.  
We observe that most of the modes show a tangible influence on the $U$ value. Remarkably, the $R_5^-$ mode experiences a change of 113 cm$^{-1}$. 
Interestingly, when the $U$ correction approaches 6.0 eV, the difference in the frequencies between the FM and $\Gamma_{5g}$ magnetic orderings, starts to decrease considerably, possibly due to an overestimation of the correlation.


\begin{figure}[b]
 \centering
 \includegraphics[width=8.7cm,keepaspectratio=true]{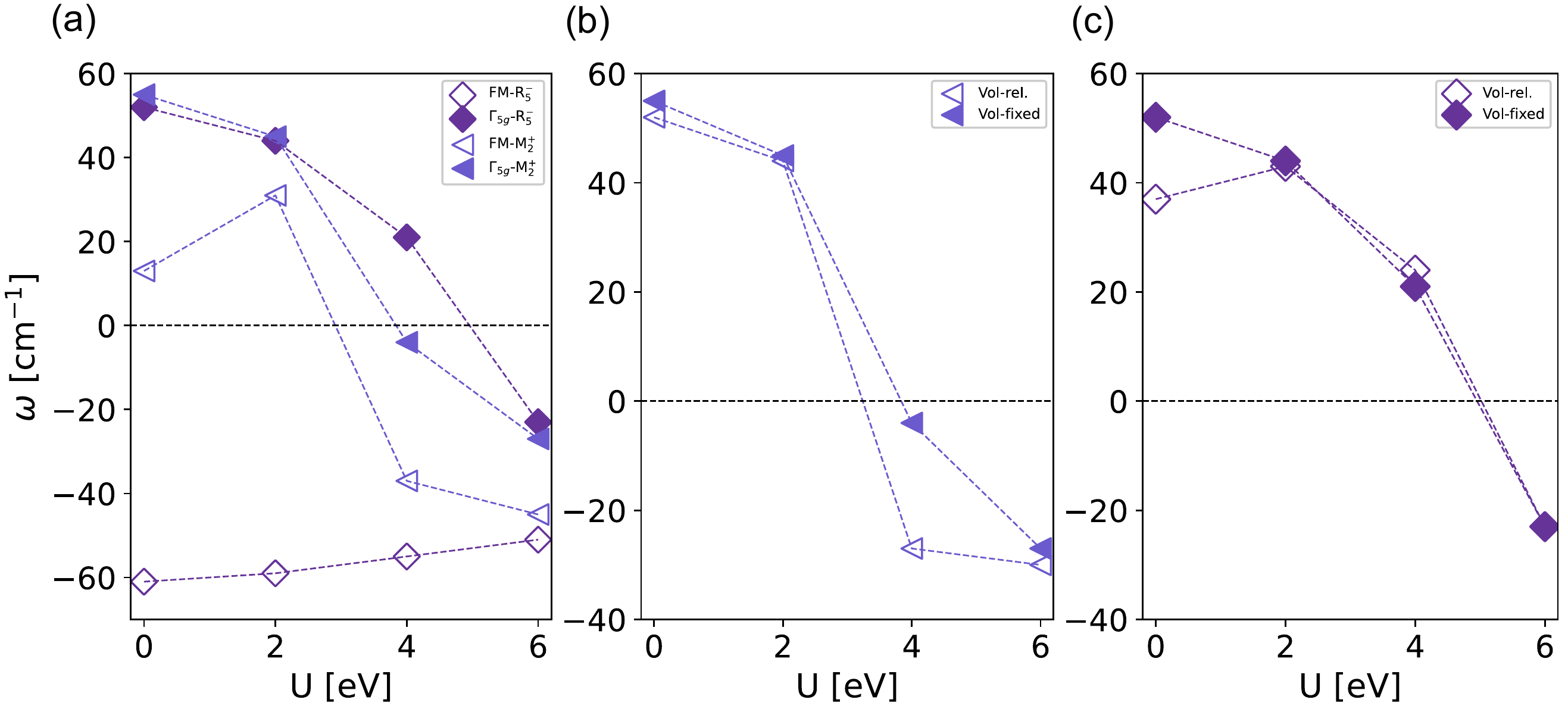}
 \caption{(Color online) In (a) are presented the $M_2^+$ and $R_5^-$ phonon frequencies computed at the  $\Gamma_{5g}$ and FM orderings in the Mn$_3$NiN for different $U$ values in which, the volume relaxation is allowed in the optimization process. Evolution of the  $M_2^+$ and  $R_5^-$ frequencies, in (c) and (c) respectively as a function of the Hubbard $U$ value. The $M_2^+$  ($R_5^-$) mode is unstable above $U$= 4.0 eV ($U$ = 6.0 eV) for both, relaxed and fixed volume, cases.}
 \label{F3}
\end{figure}


Aiming to explore the effect of the on-site Coulomb correction into the volume, and consequently, into the vibrational landscape of Mn$_3$NiN, we proceeded with the analysis by allowing the volume to relax for the different $U$ values and extracting the phonon-dispersion curves at each relaxed volume, following the procedure explained above.
Similarly, the phonons are fully stable for a small regime of $U$ values in the $\Gamma_{5g}$ and only the $M_2^+$ and $R_5^-$ modes are strongly affected by the correlation and volume increase. 
In Fig. \ref{F3}(a) we present the extracted phonon frequencies for the $R_5^-$ and $M_2^+$ modes computed within the \textsc{fm} and $\Gamma_{5g}$ magnetic orderings.
As it can be observed, the $R_5^-$ is unstable for the entire regime of $U$ values when the ferromagnetic ordering is fixed. 
The latter is corroborating that the \textsc{fm} ordering is unstable, independently of the correlation value adopted, and confirms the strong AFM exchange interaction within the Kagome lattice. 
Interestingly, the $M_2^+$ becomes unstable, in the FM magnetic ordering for $U$ values above 4.0 eV.
In Fig. \ref{F3}(b), for $M_2^+$, and Fig. \ref{F3}(c), for $R_5^-$, we present the frequencies obtained for the fixed and relaxed volumes as a function of $U$ by keeping the $\Gamma_{5g}$ ordering.
Here, we noticed that despite the volume changes, the phonon modes hold a similar trend showing an unstable behavior, for both $M_2^+$ and $R_5^-$ modes, when the $U$ values are above $U$ = 4.0 eV.

After analyzing the antiperovskite structure, it becomes clear that the vibrational modes related to rigid octahedral rotations and tilting will be strongly coupled to the magnetic behavior thanks to the magnetic exchange interactions that, in the antiperovskite case, takes place in the Mn--Mn corner-to-corner interaction. 
As such, the strong magnetic frustration will tend to keep the triangular structure unaffected. However, the hypothetical  \textsc{fm} and/or external constrains such as strain, could introduce into the system changes in the electronic structure taht could have as a consequence, dislocations of the Kagome lattices formed along with the [111] family of planes thanks to the strong magnetoestructural coupling.
Therefore, we could expect that induced rotations and tiltings at experimentally growth thin-films, onto substrates with such octahedral patterns, might trigger the ferromagnetic state in the Mn$_3$NiN antiperovskite.

As observed in the literature for the perovskite oxides counterparts, the strongest spin-lattice coupling phenomenon is observed in manganite-based oxides, such as SrMnO$_3$ \cite{PhysRevB.84.104440, PhysRevB.85.054417,PhysRevLett.104.207204}. In this manganite oxide, it is possible to achieve a switching in the magnetic structure and to induce phase transitions by external constraints, as it might be obtained in the Mn$_3$NiN. 
Thus, in order to disentangle the frustrated lattice contribution and the manganese cation intrinsic tendency to exhibit a large spin-phonon coupling, we performed a computational experiment by computing the lattice parameter dependency on the on-site Coulomb correction in the Fe$_3$NiN antiperovskite compound. 
From Fig. \textcolor{blue}{3S}, in the Supplement Material, we observe that, although the Fe$_3$NiN is affected, the influence of the $U$ parameter is smaller, suggesting that the magnetic frustrations in the $A_3B$N  ($A$ = Fe, Mn, and Ni) family are crucial ingredients for the magnetostructural behavior that it is enhanced by the manganese contribution.

\section{Conclusions}
\label{conclusions}
We have investigated, by means of first-principles calculations, the magnetostructural behavior in the Mn-based nitride Mn$_3$NiN atiferromagnetic antiperovskite by exploring the spin-phonon coupling. 
According to our results, we found a rather strong correlation between the magnetic, electronic, and vibrational degrees of freedom. 
As such, a variation of the phonon frequencies, as well as the cell volume, was observed as a function of the $U$ correlation. 
Interestingly, phonon modes associated with in-phase and out-of-phase rotations, represented by the $M_2^+$ and $R_5^-$ respectively, can be tuned by the magnetic ordering in the Mn$_3$NiN.
Remarkably, we highlight the importance of the exchange-correlation treatment in the calculations showing its influence on the electronic, magnetic, and phononic structures.
We thus hope that our work will motivate further studies leading to a deeper exploration of the spin-phonon coupling in this antiferromagnetic antiperovskite compound and then, potentially achieving the control of the magnetic features by external constraints, such as epitaxial strain and/or applied magnetic field.

\section*{Acknowledgements}
\label{acknowledgements}
We thank prof. Aldo H. Romero for helpful comments and discussions.
The calculations presented in this paper were carried out using the GridUIS-2 experimental testbed, being developed under the Universidad Industrial de Santander (SC3-UIS) High Performance and Scientific Computing Centre, development action with support from UIS Vicerrectoría de Investigación y Extension (VIE-UIS) and several UIS research groups as well as other funding resources.
Additionally, we acknowledge the computational support extended to us by Laboratorio de Supercomputo del Sureste (LNS), Benemérita Universidad Autónoma de Puebla, BUAP, for performing heavy theoretical calculations.
A.C.G.C. acknowledge the grant No. 2677 entitled “Quiralidad y Ordenamiento Magnético en Sistemas Cristalinos: Estudio Teórico desde Primeros Principios” supported by the VIE – UIS. 

\bibliography{library}

\end{document}